\documentclass[a4paper,12pt]{article}
\usepackage{graphicx}
\usepackage{amsmath}
\usepackage{amssymb}
\usepackage{isotope}
\usepackage[section]{placeins} %Puts all figures under the same section
\usepackage{float}
\usepackage{url}
\usepackage{multicol}
\usepackage{multirow}
\usepackage{booktabs} % Allows the use of \toprule, \midrule and \bottomrule in tables for horizontal lines
\begin{document}
\begin{center}

{\large\bf Synergy of nuclear data systematics and proxy-SU(3) in planning future experiments in the superheavies mass region}\bigskip
 
 \footnotesize

S.K.Peroulis$^{1,2}$, S.B. Bofos$^1$, T. J. Mertzimekis$^1$, A. Martinou$^2$, D. Bonatsos$^2$

\medskip

{\it 
$^1$~Department of Physics,University of Athens, Zografou Campus, GR-15784, Athens, Greece\\
$^2$~Institute of Nuclear and Particle Physics, National Centre for Scientific Research ''Demokritos'', GR-15310 Aghia Paraskevi, Athens, Greece
}
\end{center}
\noindent
\rule{17.6cm}{0.3mm}

\bigskip

\footnotesize
\noindent {\bf Abstract:} \hskip 3mm
Spectroscopic information of hard-to-reach superheavy nuclei can be invaluable in understanding the dynamics of nuclear systems at large values of charge and volume. RIB factories of the next generation, such as FAIR, plan to provide heavy ion beams at high energies to facilitate experimental access to these mass regimes. In preparation of future experimental endeavours, a systematic survey of available nuclear data, mainly energies and reduced transition probabilities/lifetimes of short-lived 2$^+_1$ states in even-even isotopes with Z=82, 84, 86 was undertaken. The principle motivation is to trace the competition between collective and single-particle degrees of freedom in the mass area just above \isotope[]{Pb} (Z=82), an area known to exhibit isomerism, octupole degrees of freedom and shape coexistence. Existing data were compared to the theoretical predictions using the analytical, parameter-free proxy-SU(3) scheme, for neutron numbers N=96-116. The model was further employed to predict currently unknown values for spectroscopic data in series of \isotope[]{Pb}, \isotope[]{Po} and \isotope[]{Ra} isotopes.

\bigskip


\noindent
\rule{17.6cm}{0.3mm}
\pagestyle{empty}

\normalsize

\section{Introduction}
\noindent The next-generation radioactive beam facilities plan to produce exotic nuclei and use them as probes of nuclear phenomena hardly known or entirely unknown today. Shedding light on the nuclear structure in the extremes of the nuclear chart, as well as in the superheavy region, can offer new insights to understanding the dynamics of the subatomic world. Preparing for the future, proper synergies between available experimental data and theoretical modeling have to be built to lay the ground for important discoveries. In this work, we report on a survey of spectroscopic data in the superheavies and the application of a recently developed model, proxy-SU(3), to provide predictions for the same set.

\section{Data Survey}
\noindent Existing experimental values of  the first $2^{+}$ energy state in the ground state band E(2$^+_1$) and halflives $\tau$ were surveyed \cite{nndc,exp}. Using the above values, B(E2) values are extracted according to the simplified expression:

\begin{equation}\label{eq:be2_exp}
B ( E2 \downarrow ) = 0.0816 \frac { 1 } { \tau {\left( E_{2_{1^+}} \right)}^{5}}
\end{equation}

\section{The proxy-SU(3) scheme}

\noindent Proxy-SU(3) is a relatively new algebraic approach \cite{proxy} to describe properties of nuclei based on fermionic symmetries. The proxy-SU(3) scheme is a good approximation to the full set of orbitals in a major shell. Complex shell-model calculations have been replaced with a symmetry-based description, thus enabling the prediction of several nuclear properties \textbf{analytically} and often in a \textbf{parameter-free} way.\\

Each nucleus is characterized by its irreducible representation (irrep), noted as $(\lambda,\mu)$. Those irreps are organized in tables for every Z and/or N \cite{prolate}. This scheme has been proven to work best for heavier nuclei, precisely where full microscopic calculations are most challenged.

Despite its recent development, proxy-SU(3) has already been successful in predicting the prolate over oblate dominance in deformed nuclei, the determination of the prolate to oblate transition, and parameter free predictions for the deformation parameters $\beta$ and $\gamma$ for even-even rare earths and superheavy elements \cite{prolate,letter,martinou1}, which were successfully compared to existing theoretical models, such as Relativistic Mean Field.

\subsection{B(E2) values within the proxy-SU(3) scheme}

\noindent B(E2) values are extracted within the proxy-SU(3) scheme using the expression \cite{draayer}:

\begin{equation}\label{eq:be2_proxy}
B \left[ E 2 ; ( \lambda , \mu ) K L \rightarrow ( \lambda , \mu ) K ^ { \prime } L ^ { \prime } \right] = 4 \left( \frac { Z } { S } \right) ^ { 2 } C _ { 2 } ^ { ( \lambda , \mu ) } \frac { 2 L ^ { \prime } + 1 } { 2 L + 1 } \left\langle ( \lambda , \mu ) K L ; ( 1,1 ) 2 | | ( \lambda , \mu ) K ^ { \prime } L ^ { \prime } \right\rangle ^ { 2 }
\end{equation}

\noindent where:
\begin{itemize}
\item $( \lambda , \mu )$ is the irrep that characterizes the specific nucleus\\
\item $K$,$K^{\prime}$ are the missing quantum numbers of the initial and final state. When an $SU(3) \supset SO(3)$ decompositon appears, they are required for the separation of the energy bands.\\
\item $S = S_{\pi} + S_{\nu}$ is the size of the shell\\
\item $C _ { 2 } ^ { ( \lambda , \mu ) } = \lambda ^ { 2 } + \mu ^ { 2 } + 3 ( \lambda + \mu ) + \lambda \mu$ is the second order Casimir operator of SU(3)\\
\item $L$,$L^{\prime}$ are the angular momenta of the initial and final energy level, and finally\\
\item ${\left\langle (\lambda,\mu) K L ; (1,1) 2 || (\lambda,\mu) K^{\prime} L^{\prime} \right\rangle}^{2}$ are coefficients relevant to the decomposition from SU(3) to SO(3). Those coefficients are obtained using the \textit{SU3CGVCS} code \cite{bahri}.
\end{itemize}

\noindent Based on equation \eqref{eq:be2_proxy}, analytical calculations of B(E2) values in the rare earth region have been presented in \cite{martinou1}.

\subsection{Proxy-SU(3) and shape coexistence}

\noindent Lately, it has been observed that in areas where shape coexistence \cite{heyde} is expected, the ground state band and the nearby lying $K=0$ band are represented by the proxy-SU(3) irreps and the exact SU(3) irreps occurring by consideration of the magic numbers of the isotropic three-dimensional harmonic oscillator \cite{assimakis,martinou2}.
\newpage
\section{Results \& Discussion}
\subsection{E(2$^+_1$) Data Survey}
\begin{figure}[H]
{\includegraphics[width=90mm]{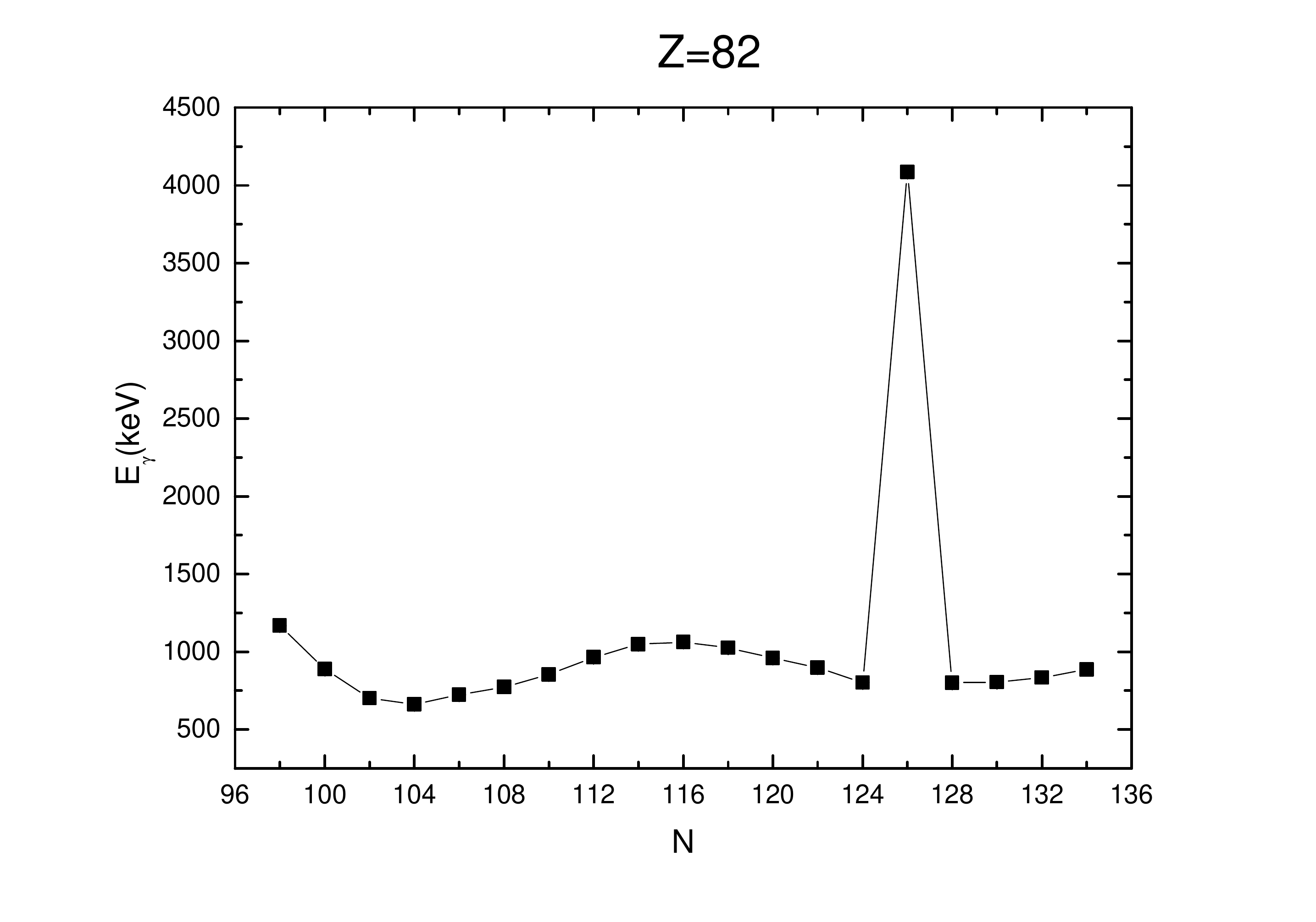}\hspace{1mm} 
\includegraphics[width=90mm]{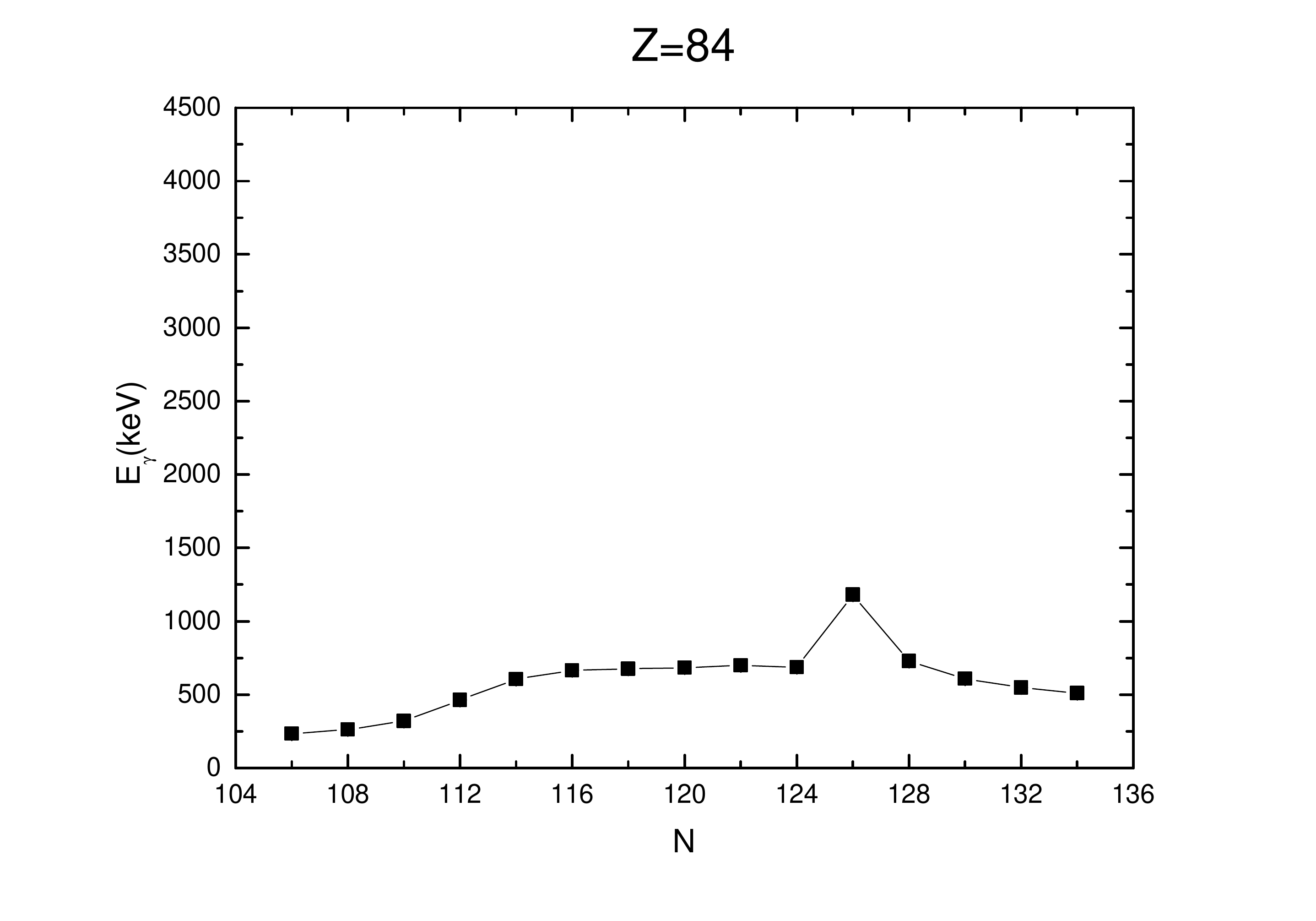}}
\begin{center}
\includegraphics[width=90mm]{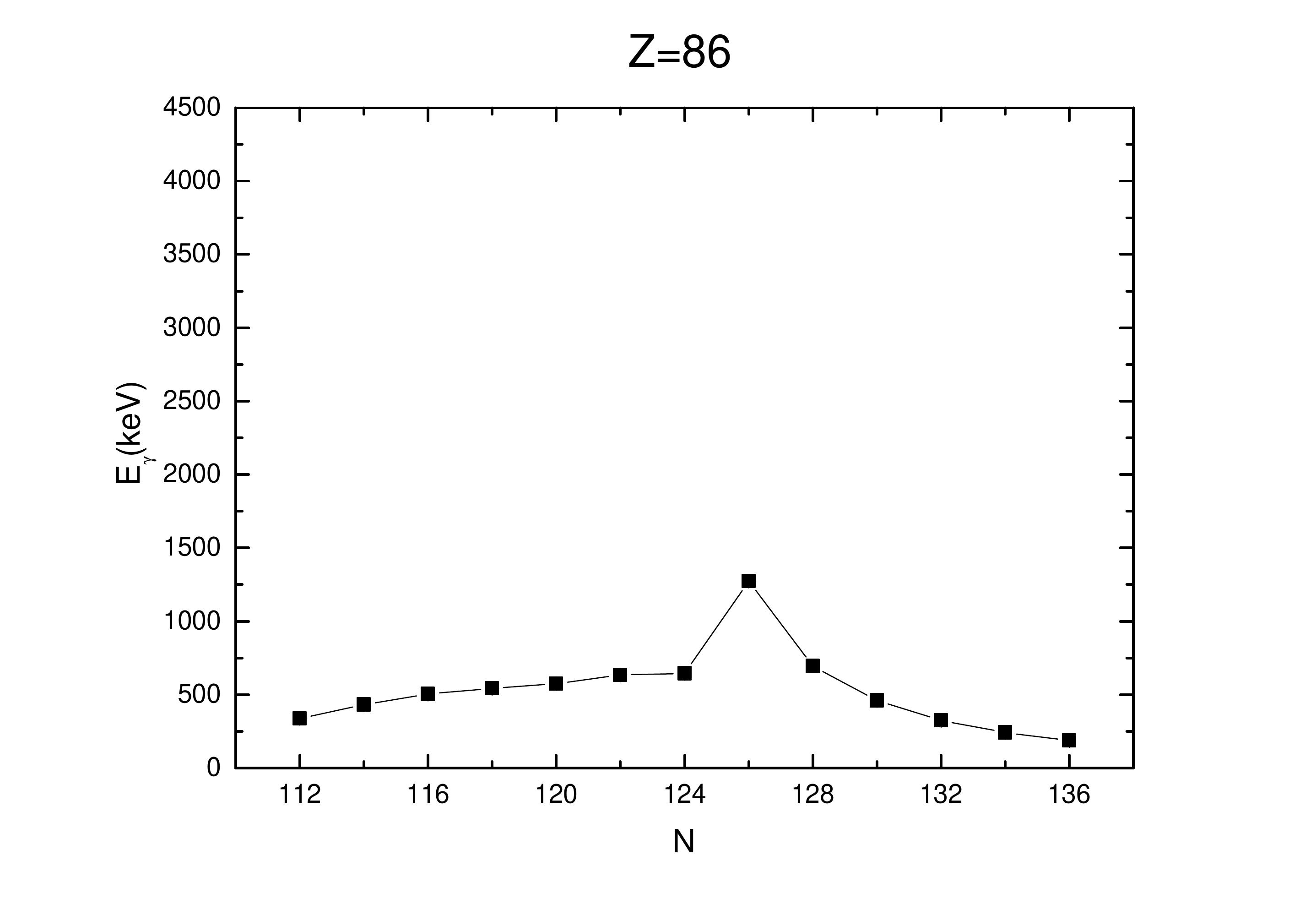}
\end{center}
\caption{Experimental values of the energies of the 2$^+_1$ states for \isotope[]{Pb}, \isotope[]{Po} and \isotope[]{Rn} isotopes.}
\label{fig:e2_plots}
\end{figure}

%\begin{figure}[h!]
%\includegraphics[width=0.3\textwidth]{Pb_E_g.pdf}
%\includegraphics[width=0.3\textwidth]{Po_E_g.pdf}
%\includegraphics[width=0.3\textwidth]{Rn_E_g.pdf}
%\caption{Experimental values of the energies of the 2$^+_1$ states for \isotope[]{Pb}, \isotope[]{Po} and \isotope[]{Rn} isotopes.}
%\label{fig:e2_plots}
%\end{figure}

\subsection{B(E2)s in Superheavies Region \& Candidate Neutron Numbers for Shape Coexistence}

In the present work we are interested in the superheavies region focusing on the  \isotope[]{Pb}, \isotope[]{Po} and \isotope[]{Rn} isotopic chains. It is necessary to stress the fact that no other analytical calculation is known in this region, mainly due to the complexity of these systems.

\begin{figure}[H]
{\includegraphics[width=90mm]{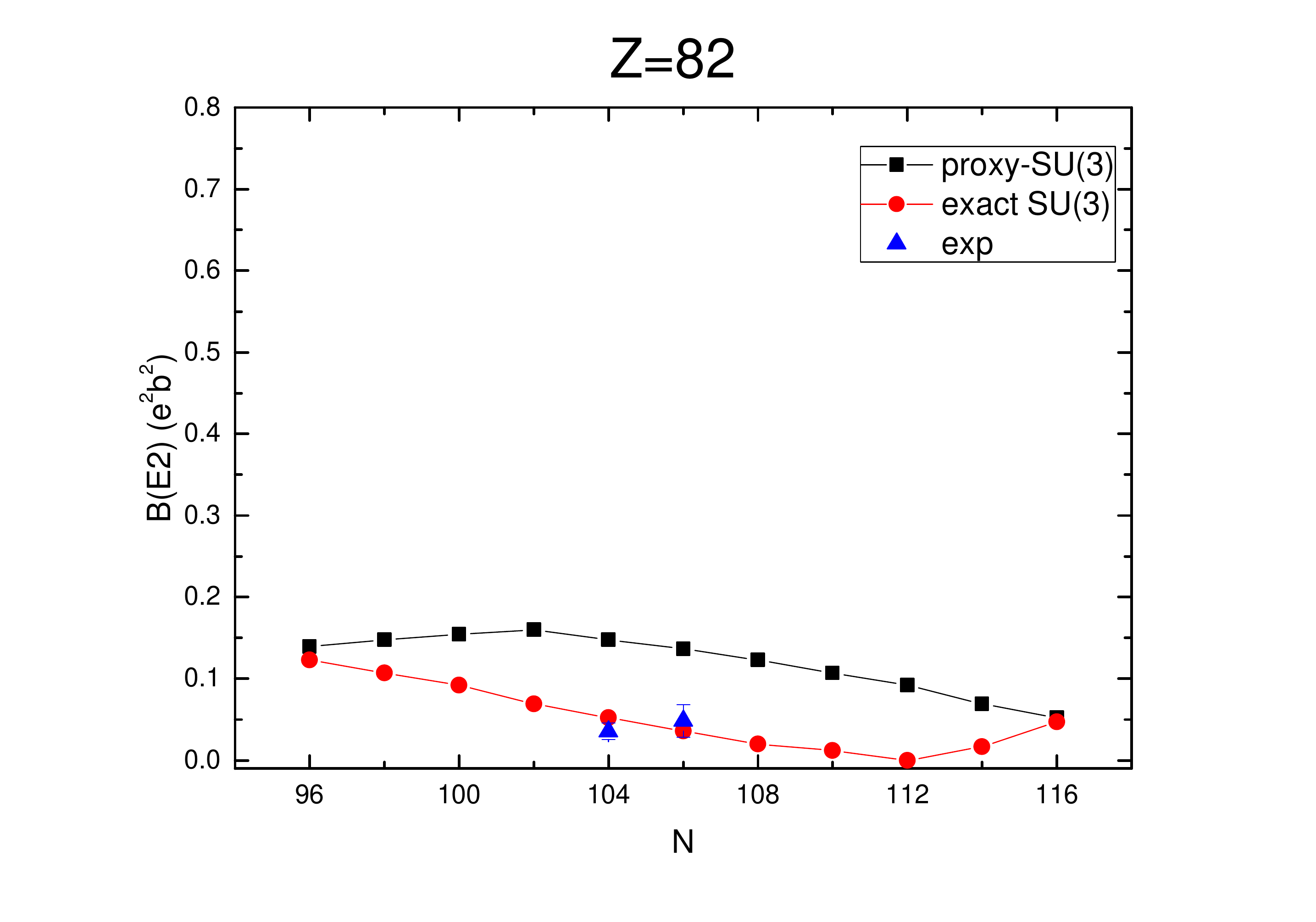}\hspace{1mm} 
\includegraphics[width=90mm]{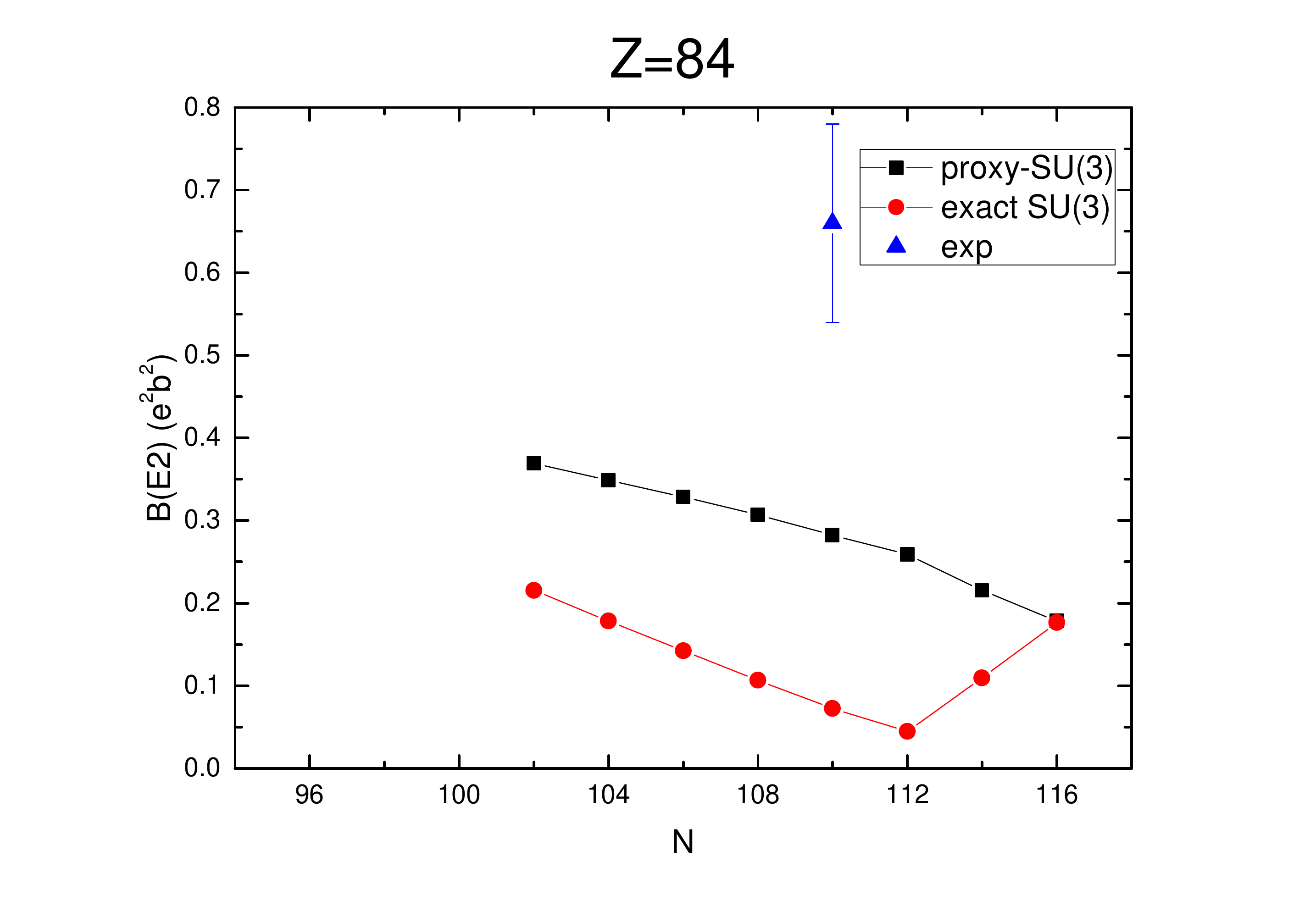}}
\begin{center}
\includegraphics[width=90mm]{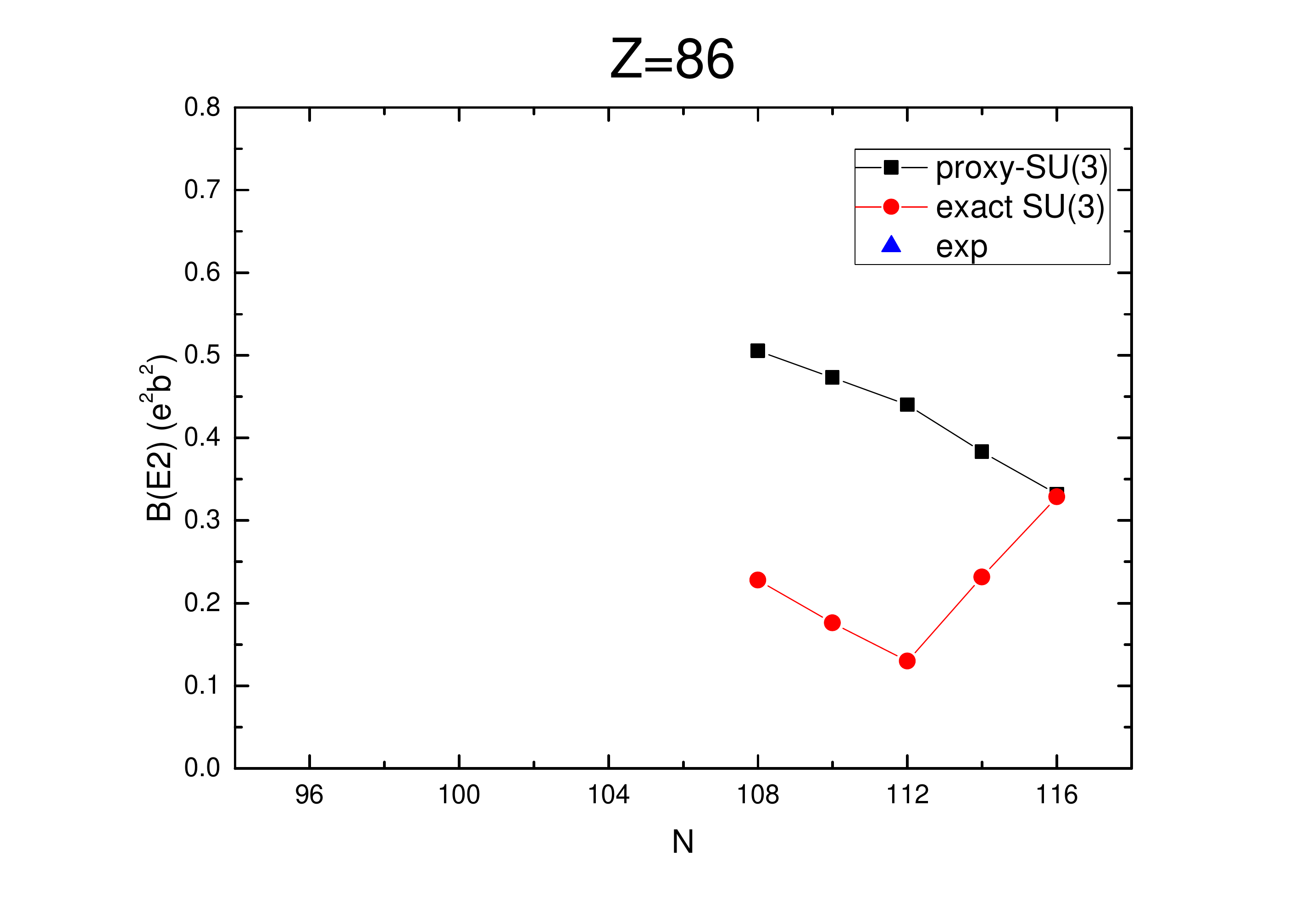}
\end{center}
\caption{Theoreticaly predicted B(E2) values for \isotope[]{Pb}, \isotope[]{Po} and \isotope[]{Rn} isotopes, using both proxy-SU(3) and exact SU(3) schemes. All known experimental data have been included for comparison.}
\label{fig:be2_plots}
\end{figure}

In Table \ref{tab:irreps}, the relative irreps (both proxy-SU(3) and exact SU(3)) are presented. Using these irreps in equation \eqref{eq:be2_proxy}, we first obtain the SU(3) $\supset$ SO(3) coefficients and then we get the B(E2) values in $e^2b^2$. Our results are presented in Fig. \ref{fig:be2_plots}.

%-----------------------------------------------------------------------
%	IRREPS Correct
%-----------------------------------------------------------------------
\begin{table}[H]
\centering
\begin{tabular}{c c c c c c c c c c}
\toprule
Element &	\multicolumn{3}{c}{\isotope[]{Pb}} &	\multicolumn{3}{c}{\isotope[]{Po}} &	\multicolumn{3}{c}{\isotope[]{Rn}}\\
\midrule
N				&		A		&		$(\lambda,\mu)_{exact}$		&		$(\lambda,\mu)_{proxy}$		&		A		&		$(\lambda,\mu)_{exact}$		&		$(\lambda,\mu)_{proxy}$		&		A		&		$(\lambda,\mu)_{exact}$		&		$(\lambda,\mu)_{proxy}$	\\
\midrule
\midrule
96	&	178	&	(28,12)	&	(34,6)		&								&						&			&						&					\\
98	&	180	&	(28,8)		&	(34,8)		&								&						&			&						&					\\
100	&	182	&	(30,0)		&	(36,6)		&								&						&			&						&					\\
102	&	184	&	(20,10)	&	(40,0)		&	186	&	(40,10)	&	(60,0)			&			&						&					\\
104	&	186	&	(12,16)	&	(34,8)		&	188	&	(32,16)	&	(54,8)			&			&						&					\\
106	&	188	&	(6,18)		&	(30,12)	&	190	&	(26,18)	&	(50,12)		&			&						&					\\
108	&	190	&	(2,16)		&	(28,12)	&	192	&	(22,16	)	&	(48,12)		&	194	&	(34,20)		&	(60,16)	\\
110	&	192	&	(0,10)		&	(28,8)		&	194	&	(20,10)	&	(48,8)			&	196	&	(32,14)		&	(60,12)	\\
112	&	194	&	(0,0)		&	(30,0)		&	196	&	(20,0)		&	(50,0)			&	198	&	(32,4)			&	(62,4)		\\
114	&	196	&	(12,0)		&	(20,10)	&	198	&	(32,0)		&	(40,10)		&	200	&	(44,4)			&	(52,14)	\\
116	&	198	&	(20,2)		&	(12,16)	&	200	&	(40,2)		&	(32,16)		&	202	&	(52,6)			&	(44,20)	\\
\bottomrule
\end{tabular}
\caption{The relative proxy-SU(3) and exact SU(3) irreps used in our calculations.}
\label{tab:irreps}
\end{table}
%-----------------------------------------------------------------------
%	IRREPS
%-----------------------------------------------------------------------
%\begin{table}[H]
%\centering
%\begin{tabular}{c c c c c}
%\toprule
%Element & N & A & $(\lambda,\mu)_{exact}$ &	$(\lambda,\mu)_{proxy}$\\
%\midrule
%\midrule
%\textbf{Pb} & 96 & 178 & $(28,12)$ & $(34,6)$\\
%~& 98 &	180 & $(28,8)$ & $(34,8)$\\
%~& 100 & 182 & $(30,0)$ & $(36,6)$\\
%~& 102 & 184 & $(20,10)$ & $(40,0)$\\
%~& 104 & 186 & $(12,16)$ & $(34,8)$\\
%~& 106 & 188 & $(6,18)$ & $(30,12)$\\
%~& 108 & 190 & $(2,16)$ & $(28,12)$\\
%~& 110 & 192 & $(0,10)$ & $(28,8)$\\
%\midrule				
%\textbf{Po} & 102 &	186	& $(40,10)$ & $(60,0)$\\
%~& 104 & 188 & $(32,16)$ & $(54,8)$\\
%~& 106 & 190 & $(26,18)$ & $(50,12)$\\
%~& 108 & 192 & $(22,16)$ & $(48,12)$\\
%~& 110 & 194 & $(20,10)$ & $(48,8)$\\
%\midrule																
%\textbf{Rn} & 108 & 194 & $(34,20)$ & $(60,16)$\\
%~& 110 & 196 & $(32,14)$ & $(60,12)$\\
%\bottomrule
%\end{tabular}
%\caption{The relative proxy-SU(3) and exact~SU(3) irreps used in our calculations.}
%\label{tab:irreps}
%\end{table}

%\begin{figure}[h!]
%\includegraphics[width=0.3\textwidth]{plot_BE2_Pb.pdf}
%\includegraphics[width=0.3\textwidth]{plot_BE2_Po.pdf}
%\includegraphics[width=0.3\textwidth]{plot_BE2_Rn.pdf}
%\caption{Theoreticaly predicted B(E2) values for \isotope[]{Pb}, \isotope[]{Po} and \isotope[]{Rn} isotopes, using both proxy-SU(3) and exact SU(3) schemes. All known experimental data have been included for comparison.}
%\label{fig:be2_plots}
%\end{figure}
\newpage
\section{Conclusions}

The above observations can be summarized as follows:

\begin{enumerate}
\item Several E(2$^+_1$) experimental values with low uncertainties are available, however, \textbf{very few} lifetimes values exist, with rather \textbf{huge} uncertainties, mainly due to the experimental difficulties of measuring half-lives in heavy nuclei.\\
\item The trend in the energies in Fig.~\ref{fig:e2_plots} (plateaus instead of constantly in/de-creasing behavior) can be considered as indication of shape coexistence besides already known collective phenomena in the area.\\
\item The exact SU(3) predictions (Fig.~\ref{fig:be2_plots}) seem to agree well with the available experimental values in \isotope[]{Pb} isotopes. However, more data are needed to examine any shape coexistence phenomena that may occur in the region.\\
\item The region above \isotope[]{Pb} is widely unexplored regarding any spectroscopic information, mainly due to experimental limitations. RIB factories of near future (e.g. FAIR, FRIB) plan to provide heavy probes at significant beam intensities aiming at exploring a widely unknown mass region.
\end{enumerate}

To conclude, a synergy between surveyed experimental data and theoretical models has been presented in this paper, identifying open questions in the superheavy mass. Such questions have to included as motivation for designing future experimental activities at RIB factories.


\begin{thebibliography}{99}

\bibitem{nndc}
National Nuclear Data Center, \url{http://www.nndc.bnl.gov}

\bibitem{exp}
T. Grahn et al., Nucl. Phys. A 801, 83 (2008)

\bibitem{proxy}
D. Bonatsos et al., Phys. Rev. C 95, 064325 (2017)

\bibitem{prolate}
D. Bonatsos et al., Phys. Rev. C 95, 064326 (2017)

\bibitem{letter}
D. Bonatsos, Eur. Phys. J. A 53 , 148 (2017)

\bibitem{martinou1}
A. Martinou et al., HNPS2017 Proceedings, arXiv:1712.04134

\bibitem{draayer}
J. P. Draayer et al., Ann. Phys. (N.Y.) 156, 41 (1984) 

\bibitem{bahri}
C. Bahri et al., Comp. Phys. Comm. 159, 121 (2004)

\bibitem{heyde}
K. Heyde and J.L.Wood, Rev. Mod. Phys. 83, 1467 (2011)

\bibitem{assimakis}
I.E. Assimakis et al., these proceedings

\bibitem{martinou2}
A. Martinou et al., these proceedings 

\end{thebibliography}
\end{document}